# Direct evidence for Cooper pairing without a spectral gap in a disordered superconductor above $T_c$

Koen M. Bastiaans[1], Damianos Chatzopoulos[1], Jian-Feng Ge[1], Doohee Cho[2], Willem O. Tromp[1], Jan M. van Ruitenbeek[1], Mark H. Fischer[3], Pieter J. de Visser[4], David J. Thoen[5,6], Eduard F.C. Driessen[7], Teunis M. Klapwijk[5,8,9], Milan P. Allan[1,*]

[1]Leiden Institute of Physics, Leiden University, Niels Bohrweg 2, 2333 CA Leiden, the Netherlands
[2]Department of Physics, Yonsei University, Seoul 03722, Republic of Korea
[3]Department of Physics, University of Zurich, Winterthurersstrasse 190, 8057 Zurich, Switzerland
[4]SRON Netherlands Institute for Space Research, Sorbonnelaan 2, 3584 CA Utrecht, the Netherlands
[5]Kavli Institute of Nanoscience, Delft University of Technology, Lorentzweg 1, 2628 CJ Delft, the Netherlands
[6]Faculty of Electrical Engineering, Mathematics and Computer Science, Delft University of Technology, Mekelweg 4, 2628 CD Delft, the Netherlands
[7]Institut de Radio Astronomie Millimétrique (IRAM), Grenoble, 300 Rue de la Piscine, 38400 Saint-Martin-d'Hères, France
[8]Institute of Topological Materials, Julius-Maximilian-Universität Würzburg, Sanderring 2, 97070 Würzburg, Germany
[9]Physics Department, Moscow State University of Education, 1/1 Malaya Pirogovskaya Str., 119991 Moscow, Russia

*Correspondence to: allan@physics.leidenuniv.nl.

*The idea that preformed Cooper pairs could exist in a superconductor above its zero-resistance state has been explored for unconventional, interface, and disordered superconductors, yet direct experimental evidence is lacking. Here, we use scanning tunneling noise spectroscopy to unambiguously show that preformed Cooper pairs exist up to temperatures much higher than the zero-resistance critical temperature $T_c$ in the disordered superconductor titanium nitride, by observing a clear enhancement in the shot noise that is equivalent to a change of the effective charge from 1 to 2 electron charges. We further show that spectroscopic gap fills up rather than closes when increasing temperature. Our results thus demonstrate the existence of a novel state above $T_c$ that, much like an ordinary metal, has no (pseudo)gap, but carries charge via paired electrons.*

Disordered superconductors, like the titanium nitride (TiN) thin films that are the focus of this study, are close to a superconductor-metal or superconductor-insulator transition *(1)*, and often exhibit electronic granularity, either emergent *(2)* or due to small superconducting islands coupled to each other. Above $T_c$, disordered superconductors exhibit unusual normal state properties, often including anomalous resistivity *(3-7)* and a so-called pseudogap in spectroscopic properties *(8)*. Similar to the case of high-temperature *(9)*, interface *(10,11)*, and heavy fermion superconductors *(12)*, these properties have sometimes been interpreted as signatures of short-lived, fluctuating *(5,13)*, or pre-formed Cooper pairs *(1,14)* that do not form a phase-coherent state. Such a situation would be in strong contrast to conventional, elemental superconductors, where pairing and condensation take place concurrently at the critical temperature $T_c$.

The concept of pairing without phase coherence (Fig. 1a) was introduced even before the famed Bardeen-Cooper-Schrieffer theory explained the microscopic origin of superconductivity *(15)*, and new theoretical models are still being explored. The most well-known models, going back to Fisher *(16)* and others, postulate a phase-fluctuation driven breakdown of coherence at $T_c$. The Berezinskii–Kosterlitz–Thouless formalism brings a particularly intuitive picture of such a transition: fluctuating vortex-antivortex pairs that exist below $T_c$, unbind at $T_c$, leading to the change from superconducting to resistive state as the temperature is increased *(17,18)*. Phase fluctuations are also at the core of models involving the Bose-Einstein condensation to Bardeen-Cooper-Schrieffer (BEC-BCS) crossover *(19)* which were realized in cold-atom ensembles. An alternative possibility to break down superconductivity is a decrease of the order parameter amplitude, caused by enhanced Coulomb repulsion in disordered systems *(20)*. The former models invoke paired electrons above $T_c$, while the latter does not.

The challenge to experimentally discern phase-incoherently paired electrons from single electrons is twofold. First, many properties of paired but uncondensed electrons are the same as for single electrons, including the charge transported per electron. Second, spectroscopic signatures of paired electrons are often similar to single-electron phenomena like charge density waves *(9)*. Still, a large number of intriguing observations connected to pairing above $T_c$ have been reported. Kinks in resistivity versus temperature curves and deviations from assumed normal state resistances have been connected to pairing fluctuations *(21)*. The Nernst effect has shown unusual signatures in several materials, compatible with short-lived Cooper-pair fluctuations *(22,23)*. In underdoped cuprates, enhanced noise signatures have further been reported in planar junctions and interpreted as multiple Andreev reflections enabled by pairing of charge carriers above $T_c$ *(24)*. Several spectroscopic techniques show (partially filled) gaps in the spectral weight at the Fermi level, frequently called pseudogaps, that persist above $T_c$. Taken together, the observations described here point towards a transition at $T_c$ from a macroscopic quantum state with zero resistance, towards a state where the resistance becomes finite, but which is in many respects different from the conventional metallic state of ordinary metals.

Thus motivated, our study aims to determine the nature of the charge carriers in this unconventional normal state in disordered superconductors by focusing on the effective charge of the carriers in tunneling experiments, measured through noise spectroscopy. Shot noise spectroscopy in mesoscopic systems has proven to be a powerful tool to determine the effective charge, e.g. in superconductors or in fractional quantum Hall experiments *(25)*. In general, tunneling between two leads biased with voltage $V$ is a Poissonian process. The current noise $S$ associated with the granularity of charge is proportional to the effective charge $q^*$ of the carrier and the current $I$, i.e. $S=2q^*I$. This relation allows to extract the effective charge of the carriers, which, in metal-insulator-superconductor interfaces (NIS), is equal to one electron charge ($1e$) at biases above the superconducting gap (Fig. 1b), but $2e$ within the gap. The latter is a result of Andreev reflections from paired electrons which double the effective charge transported *(26)* as illustrated in Fig. 1c. The signature for paired electrons is thus simple and unambiguous: in a tunneling experiment from a normal metal to a system with bound pairs, the normalized noise should change from $S/2I=1e$ to $S/2I=2e$ when the bias is reduced to below the gap energy. Experiments involving conventional superconductors have confirmed the doubling, and even further multiplication, of shot noise as a tell-tale signature of paired electrons *(26-29)*, but ensuring a clean vacuum barrier has shown to be challenging.

We choose to use the disordered superconductor titanium nitride *(30)* for this study as it exhibits robust signatures of unusual physics above $T_c$ without any competing orders such as charge density waves *(9)*. Our TiN films develop a zero-resistance state, i.e. become superconducting below 2.95 K as determined by transport measurements, and exhibit a mean free path of 0.57 nm, with a coherence length of ~10 nm, placing them well within the so-called 'dirty limit' of superconductivity. We use 45 nm thick films fabricated by plasma-enhanced atomic layer deposition (ALD) *(30)*, and as control samples 60 nm thick films made by sputter deposition on silicon substrates (for data from the sputtered sample is qualitatively similar and shown in the Methods). The samples are inserted into an ultrahigh vacuum chamber, and inserted into a cryogenic scanning tunneling microscope (STM). The noise measurements are done with a custom-built, cryogenic MHz amplifier, consisting of a superconducting tank circuit and custom-built high electron mobility transistors (HEMT), as described elsewhere *(31)*. From the noise spectroscopic measurements, we measure the current fluctuations around a center frequency of 3 MHz to avoid mechanical resonances and unwanted $1/f$ noise, and we ensure that the vacuum tunneling barrier is clean by repeatedly measuring topographies with the STM *(32)* (see Methods). Spectroscopic imaging STM at 2.3K reveals a partially filled gap of $\Delta$=1.78 meV (with spatial variations of around 36% (see Methods).

Local tunneling noise spectroscopy is the key technique used in this study. We perform our experiments at fixed junction resistance ($R_J$), and thus the noise is expected to be proportional not only to the current, but also to the bias voltage, $S(q^*,V) = 2q^*I = 2q^*V/R_J$. At finite temperature ($T$) and low junction transmission, the formula is modified to $S(q^*,V) = 2q^*(V/R_J)\coth(q^*V/2k_BT)$, where $k_B$ is the Boltzmann constant. We extract the effective charge $q^*$ by numerically solving this formula for the observed shot-noise at each bias (see Methods for details) *(26)*. As expected, the effective charge at a bias higher than $\Delta$ is equal to one electron charge, $q^*=1e$, as shown in Fig. 2a and b. However, a clear change in the effective charge from $q^*=1e$ to $q^*=2e$ is visible in the data for voltages below gap energy. This is unambiguous evidence that the electrons in TiN films are paired below an energy of roughly ~$\Delta$ (indicated in Fig. 2 by blue shading). The reason that the noise does not rise immediately at $\Delta$, but at energies just below $\Delta$ is due to thermal broadening (see Methods for details). The shape and values of our noise spectra enable us to directly deduce pairing as the source of the noise, as opposed to fluctuating orders that might be present in the sample.

Remarkably, the noise enhancement to 2e persists when warming the sample to temperatures above the zero-resistance $T_c$. Fig. 3 shows noise spectra acquired at different temperatures ranging from 2.3 K to 7.2 K, which correspond to $0.78T_c$ and $2.43T_c$. Up to more than twice $T_c$, the noise spectra still show enhanced noise corresponding to 2e; only at $T = 2.43T_c$ does the noise become less. Given that $T_c$ is far below the temperature at which the noise is enhanced, there is another transition temperature – we denote it here by $T_p$ – associated with pairing. In a fluctuation picture, this would be the temperature at which the gap opens. Figure 4 summarizes the temperature evolution of the noise.

We can compare how the new temperature scales with the unusual transport properties that have been analyzed for a wide range of disordered superconductors. Fig. 4c shows a resistance versus temperature curve acquired in the same sample, showing the superconducting transition at $T_c$ = 2.95 K. Above $T_c$, the resistance curve shows a so-called *N*-shaped curvature (see Methods) as is typical for disordered superconductors not too close to the superconductor-insulator transition[8], and for cuprate high-temperature superconductors *(33)*. Around 11 K the resistivity passes a local maximum before it drops to zero below $T_c$, a signature that has been interpreted as the onset of superconducting fluctuations *(4,34)*. For the ALD sample, this is also roughly the temperature where the gap is expected to close if the ratio $2\Delta(0)/k_BT_p$ is given by the BCS-value of 3.52, while for the sputtered sample, it is lower.

Our most surprising experimental observation is that we observe pairing above $T_c$ even in the absence of a spectroscopic (pseudo)gap. As shown in Fig. 4a,b, the gap in the differential conductance of TiN does not close at $T_c$ when increasing the temperature, but fills up instead, i.e. $\Delta(T)$ is constant while the spectral weight inside the gap is filling up. While the measured temperature evolution of $\Delta(T)$ resembles findings in other disordered superconductors, the gap fills faster in our samples upon increasing temperature such that the gap is fully filled at $T_c$. Partial gap filling has been observed with various probes in other disordered and unconventional superconductors *(8,35–37)*. The phenomena can be calculated within models involving strong fluctuations of the order parameter *(38)* or significant level spacing in grains *(39)*, alternatively, one can postulate a significant fraction of unpaired electrons, or electrons with very small superconducting gaps, to exist in parallel to the superfluid *(35,40)*. Our data clearly shows that the current noise continues to correspond to ~2e at elevated temperatures, despite the filling of the gap. The state above $T_c$ thus behaves like an ordinary metal from a spectroscopy point of view, but with tunneling current fluctuations that indicate pairing. This is uniquely visible in shot noise experiments. Therefore, a putative coexistence of paired and unpaired electrons, as predicted to exist by theories of short-lived Cooper pairs *(1,22,23)*, is not present in TiN.

We note that the combined observation of both a filled gap and 2*e*-noise cannot be described in the same way as the well-known case of subgap current in break junctions of elemental superconductors *(26)*. In break junctions with low transparencies, the conductance inside the gap is much smaller because Andreev reflections happen with a probability proportional to the square of the transparency of the junction, $t^2$, while single electron tunneling outside the gap occurs with probability *t*. In contrast, in TiN we measure a similar conductance inside and outside that gap, independent of the junction resistance, despite the fact that our transparency is around $2.6 \times 10^{-3} \ll 1$. The probability for charge transfer in the range of 2*e*-noise is therefore still linear in transparency. Such a situation could, in principle, arise when the bunching of the probability for subsequent electron transfers is modulated because of Andreev processes within the sample, or in specific cases involving disordered metals *(26)*. More likely, a theory involving the spatial heterogeneity and correlations that are typical for disordered superconductors is needed to understand this peculiar state.

In summary, we have used local noise spectroscopy as an unambiguous probe of pairing in a disordered superconductor. We have shown that, (i) pairing dominates up to a temperature scale $T_p$ much larger than $T_c$, (ii) the energy of pairing is related to the gap energy, and (iii) even though the spectral gap is partly or fully filled, almost all observed electrons are paired, differentiating between proposals for pairing above $T_c$. Hence, we have observed a state that exhibits 2*e* noise despite having the characteristics of an ordinary metal in differential conductance, without a spectroscopic (pseudo)gap. Further, our results contradict theories of the breakdown of superconductivity that involve a large fraction of unpaired electrons.

**Acknowledgments:** We acknowledge C.W.J. Beenakker, G. Blatter, J.C. Davis, R. Fermin, J. Jesudasan, T. Mechielsen, P. Raychaudhuri, B. Sacépé, D. Scholma, N. Trivedi, and J. Zaanen for valuable discussions; **Funding:** This work was supported by the European Research Council (ERC StG SpinMelt) and by the Netherlands Organization for Scientific Research (NWO/OCW), as part of the Frontiers of Nanoscience program, as well as through a Vidi grant (680-47-536). D. Cho was supported by the National Research Foundation of Korea (NRF) grant funded by the Korea government (MSIT) (No. 2020R1C1C1007895 and 2017R1A5A1014862) and the Yonsei University Research Fund of 2019-22-0209. P.J.dV. was financially supported by the Netherlands Organization for Scientific Research NWO (Veni Grant No. 639.041.750); **Author contributions:** K.M.B, D.Cha., J-F.G., D.Cho., and W.O.T. performed the experiments and analyzed the data. P.J.dV., D.J.T., E.F.C.D. and T.M.K. fabricated and characterized the samples. All authors contributed to the interpretation of the data and writing of the manuscript. M.P.A. supervised the project.; **Competing interests:** Authors declare no competing interests.; and **Data and materials availability:** The data used in this work is available from the corresponding author upon reasonable request.

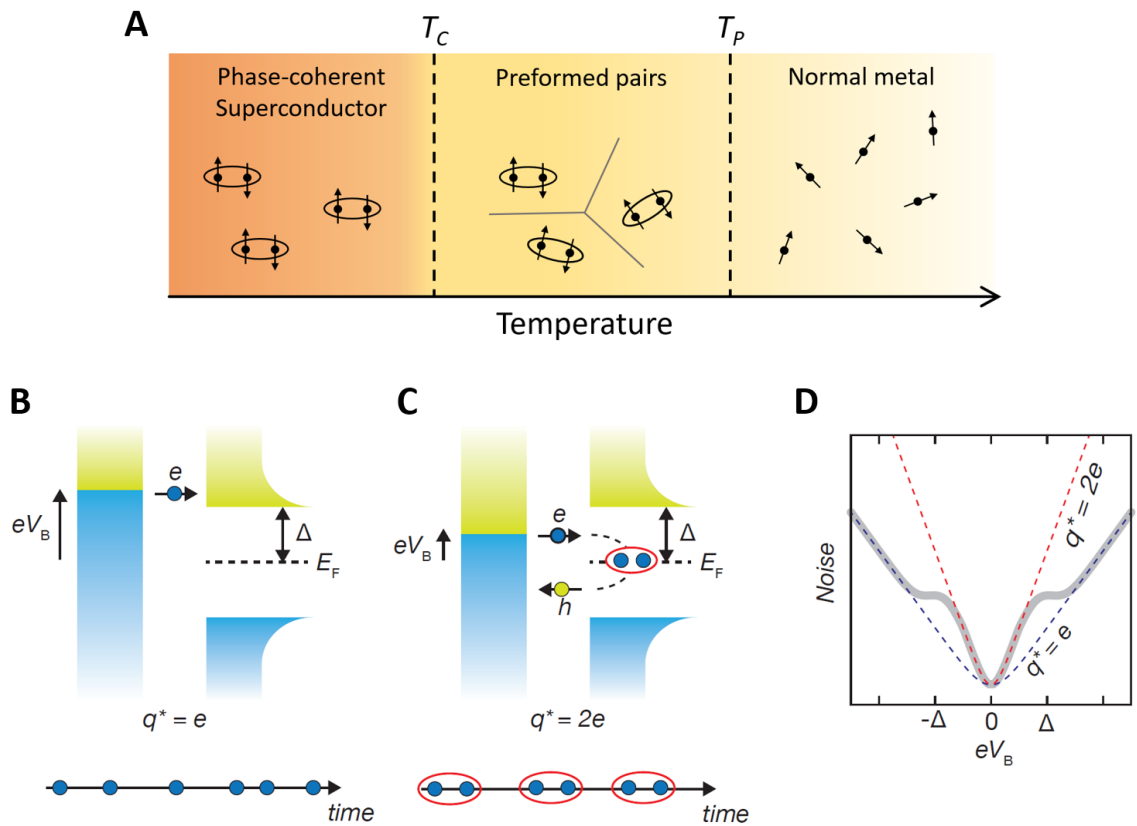

**Fig. 1.** Noise spectroscopy as a direct probe to detect paired electrons. (A). Illustration of the different electronic states. At high temperature, a conventional metal state consists of single electrons. Below Tc, these electrons couple to form a phase-coherent state of Cooper pairs. Between these two, an additional state of non-phase-coherent, preformed Cooper pairs is conjectured to exist. (B). 'Normal' NIS transport of single electron charge. The characteristic density of states of the superconducting sample is shown, with filled and empty states denoted by blue and yellow separated by a pair-breaking gap $\Delta$. (C). Andreev reflection process in a BCS superconductor. An electron transfers a Cooper pair into the superconductor by reflecting a hole in the opposite direction, effectively transferring $2e$ charge. (D). Noise as function of bias voltage for $q=1e$ and $q=2e$ transport. For a NIS junction the expected noise is indicated by the gray curve.

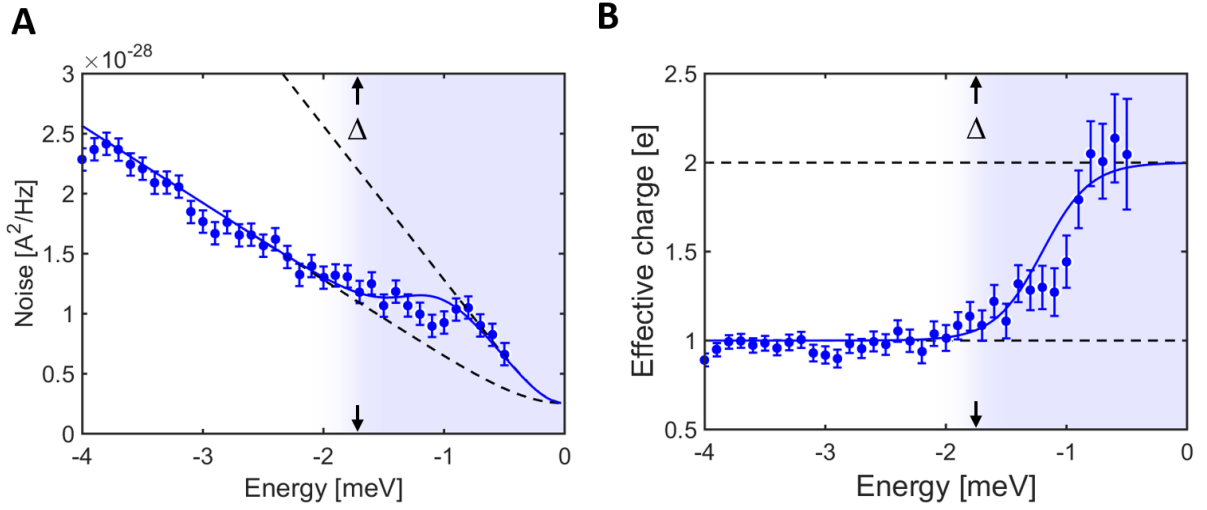

**Fig. 2.** Evidence for pairing in TiN from scanning noise spectroscopy. (A). The noise (blue dots) in the tunneling junction ($R_J$ = 5 MΩ) between the STM tip and TiN sample at 2.3 K with the thermal amplifier noise subtracted, as function of the bias voltage. Dashed lines indicate the expected noise for q=1$e$ and q=2$e$. Blue shading indicates the spectral gap observed in the differential conductance (Fig. 4a). The blue curves represents a guide to the eye to indicate thermal broadening similar to expectations from a random scattering matrix simulation (see methods). (B). Spectroscopy of the effective charge $q^*(V)$ at 2.3 K.

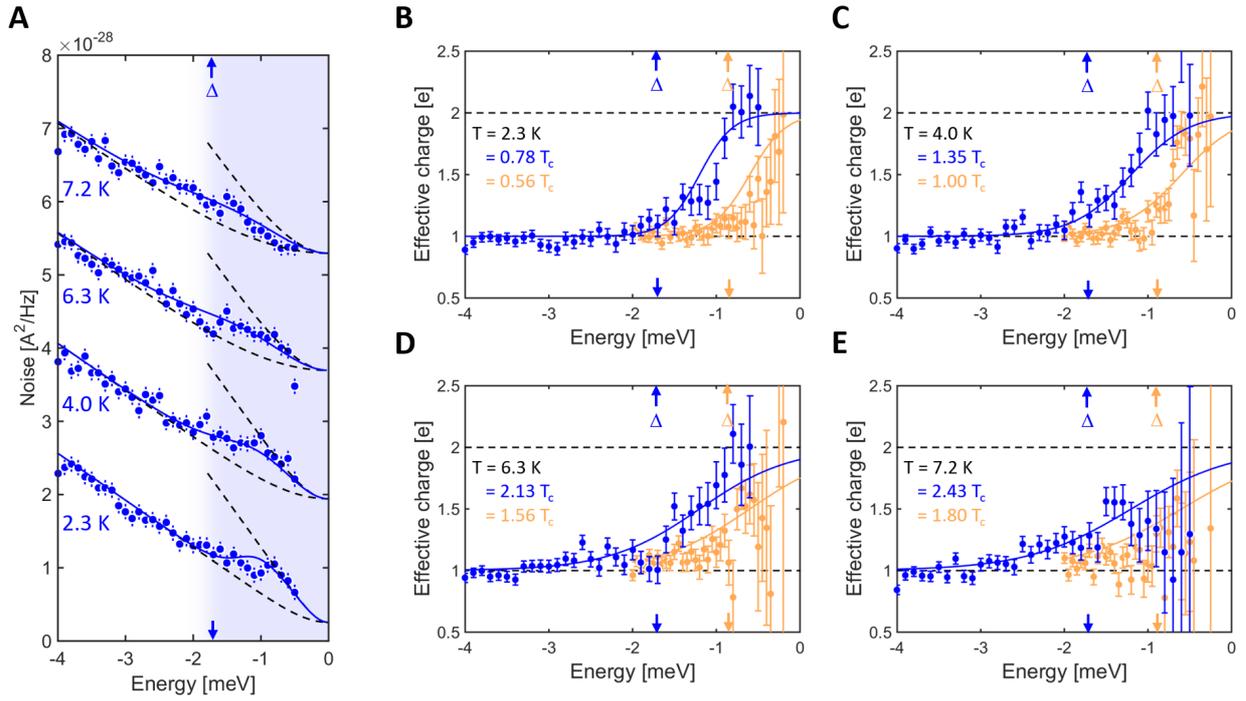

**Fig. 3.** Enhanced noise above Tc. (A). Noise spectroscopy on TiN sample for different temperature from 2.3 K = 0.78 $T_c$ to 7.2 K = 2.43 $T_c$. Blue dots indicate the measured excess noise in the junction ($R_J$ = 5 MΩ) as function of bias voltage. The different temperature curves are offset for clarity. Dashed lines indicate the expected noise for $q^*$=1e and $q^*$=2e. Blue shading highlights the spectral gap in the differential conductance. (B-E). Effective charge $q^*(V)$ for the four different temperatures. Data for the ALD (sputtered) sample is in blue (orange).

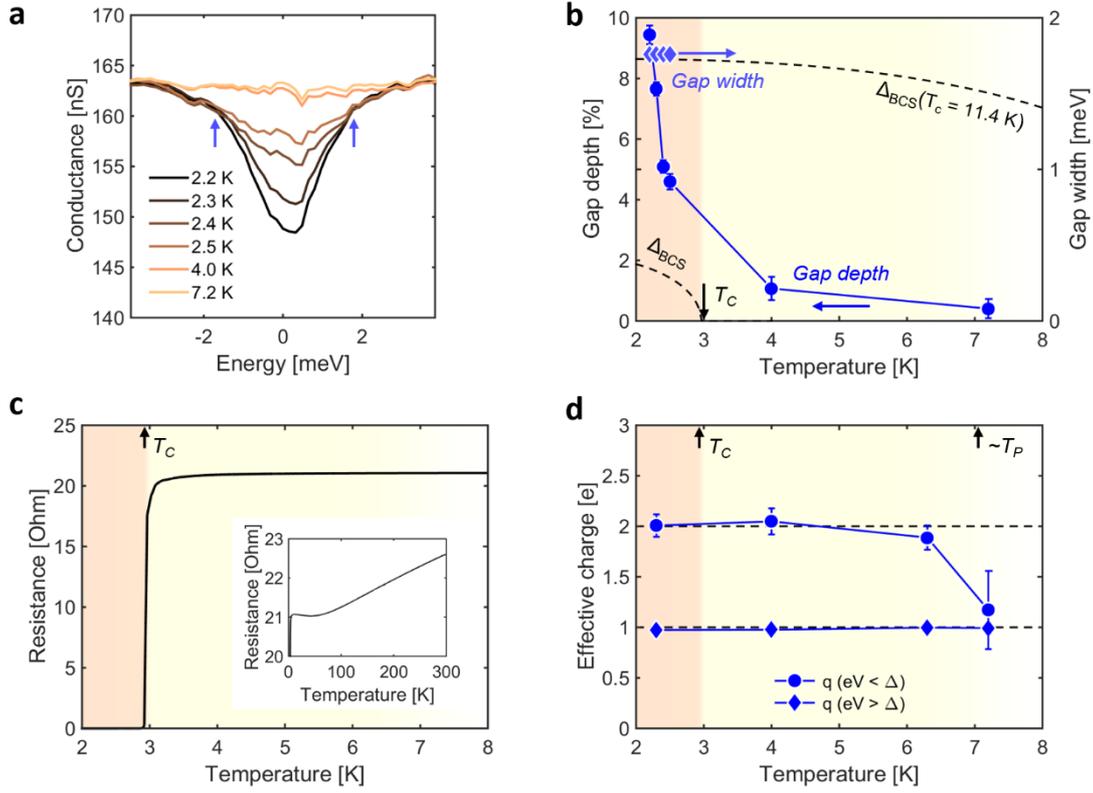

**Fig. 4.** Evidence for a preformed-pair phase above $T_c$. (A). Temperature dependence of the spectral density gap measured by the differential tunneling conductance between 2.2 K (0.74 $T_c$) and 7.2 K (2.43 $T_c$). Blue arrows indicate the gap width at 2.2 K determined by finding the minimum of second the derivative. Setup conditions: $V_{bias}$ = 5 mV, $I_{set}$ = 1 nA. (B). Gap width (blue diamonds) as a function of temperature for the curves in panel a. The dashed curves indicates the mean-field prediction for $\Delta(T_c = 2.96$ K$)$ and $\Delta(T_c = 11.4$ K$)$ from BCS theory. The depth of the gap at zero bias (blue dots) for the curves in panel a is shown in percentages with respect to the conductance at energies outside the gap. Conventional Andreev processes or thermal broadening cannot account for the filling observed here (see Methods). (C). Resistance versus temperature curve of our ALD TiN sample. The orange shaded region indicates the phase-coherent superconducting phase below the transition temperature. Inset, the resistance-temperature relation up to 300 K. (D). Effective charge outside (diamonds) and inside (circles) the spectral gap as function of temperature. The region consisting of preformed pairs includes temperatures where the gap is fully filled and is indicated by the yellow shading.

## Materials and Methods

### STM measurement schematics and topography of TiN surface

Figure S1 shows a measurement schematic and the topography of the TiN samples that we studied. TiN samples grown on Si substrates are glued using silver epoxy on a STM sample holder. The top TiN layer is electrically contacted by using silver epoxy as well, as illustrated in Fig S1a. Two- and three-dimensional topographic images of the scanned TiN surface are shown in Fig. S1 b and c, respectively. The observed topographic height variations are consistent with previously reported ones *(41,42)*.

### Deposition of the Atomic Layer Deposited (ALD) TiN film

The 45 nm thick ALD titanium nitride film was grown on a Si substrate using plasma-assisted atomic layer deposition in an Oxford Instruments reaction chamber. The precursor $TiCl_4$ and a plasma of $H_2$ and $N_2$ are stepwise repetitively introduced and react into TiN and gaseous HCl, which is a self-limiting process. After each step, the residual gas is removed from the reaction chamber by an argon gas purge. During the deposition, the substrate is heated to 400 °C and the plasma power is 400 W. More details on this sample can be found in Ref. *(30)*.

### Deposition of the sputtered TiN film

The sputtered film is deposited by reactive sputtering in a Trikon Sigma DC magnetron reactor. The 60 nm thick film is sputtered from a titanium target in a nitrogen/argon atmosphere (70/20 sccm) at room temperature and with a sputtering power of 6 kW onto a Si substrate, which is cleaned with an HF dip prior to deposition.

### Extraction of $q^*$

The noise measurements are performed with a custom-built cryogenic MHz amplifier involving an LC circuit and a HEMT that converts the current fluctuations in the junction into voltage fluctuations into a 50 Ohm line, as described in full detail elsewhere *(31)*. To extract the charge transferred in the junction we follow a similar procedure as described in in detail ref *(29)*, where we measured the effective charge in a Pb-Pb tunneling junction. The total measured voltage noise over the 50 Ohm input resistance of our Zurich Instruments MFLI Spectrum Analyzer is $S_V^m(\omega, V) = G^2 |Z_{res}^2| S_I$. Here $G$ is the total gain of the amplification chain, which we calibrated on a Pb(111) surface at all temperatures referenced in the manuscript. $Z_{res}$ is the impedance of the resonating circuit and $S_I$ is the total current noise in the circuit, consisting of three terms:

$$S_I(I) = 2q^* I \coth\left(\frac{q^* V}{2kT}\right) + \frac{4kT}{|Z_{res}|} + S_{amp},$$

where $q^*$ is the effective charge, $T$ the temperature, $k$ the Boltzmann constant and $S_{amp}$ the input noise of the amplifier. The first term represents the full junction noise and the second term the thermal noise of the LC tank. We use the full formula for the shot noise in the junction $S = 2q^* I \coth\left(\frac{q^* V}{2kT}\right)$ to numerically extract the effective charge $q^*$ from the measured noise in the following way: First, we measure the noise $S_I(0)$ with the tip retracted to obtain $\frac{4kT}{|Z_{res}|} + S_{amp}$, which is then subtracted from the noise measured in tunneling to obtain $S_I(I) - \frac{4kT}{|Z_{res}|} - S_{amp} = 2q^* I \coth\left(\frac{q^* V}{2kT}\right)$. This quantity is plotted in Fig. 2a and 3a of the main text for the ALD sample and in Fig. S6 for the sputtered sample. Next, we extract the effective charge $q^*(V,T)$ for each $V$ and $T$ numerically by finding the least square of $\left[\left(S_I(I) - \frac{4kT}{|Z_{res}|} - S_{amp}\right) - \frac{2q^* V}{R_J} \coth\left(\frac{q^* V}{2kT}\right)\right]^2$ with respect to $q^*$, plotted in Fig. 2b and Fig. 3b-e in the main text. The effective charge plotted in Fig. 4d is obtained by averaging the numerical value found for $q^*(V,T)$ over a small bias window (0.4 meV for the ALD sample, 0.2 meV for the sputtered sample) above the gap (-3.5 – -3.1 meV for the ALD sample, -1.8 – -1.6 meV for the sputtered sample) and below the gap (-0.9 – -0.50 meV for the ALD sample, -0.4 – -0.2 meV for the sputtered sample) for each temperature $T$. Points within this bias window that did not provide a solution by the least square minimization of above equation were excluded from the averaging and further error propagation (Fig. 2b and Fig. 3b-e).

Errorbars on excess noise and effective charge figures
Errorbars on the excess noise figures are determined by recording the fluctuations of the measured noise in time, and by estimating temperature fluctuations that have an influence on the thermal noise of the junction. The fluctuations of the measured noise in time are determined by time traces of the noise at constant bias at base temperature (Fig. S2, measured noise at $R_J$=2.5 MΩ and $T$=2.5 K for two different biases in tunneling and tip retracted (0 meV)). We then extract the standard deviation of these time traces and use it as the errorbar for our noise measurements, Fig. 2a and Fig. 3a. The temperature fluctuations are determined by logging the temperature at 7.2K over a few days. We use the procedure described above (Extraction of $q^*$) in order to propagate the errors of the excess noise to the effective charge versus energy plots.

Transport measurements
Transport measurements were performed on TiN samples (4.5 mm x 2.5 mm) glued on a chip carrier using Epotek H20E silver-epoxy and wired bonded to the chip carrier to make a four-probe contact. Resistance versus temperature $R(T)$ (Fig. 4c) curves are measured using the four-probe method in a pulse tube cryostat with a base temperature of 1.5 K.
The $R(T)$ curve (Fig. S3) shows a so-called $N$-shaped curvature as it is typical for disordered superconductors not too close to the superconductor-insulator transition, as well as for some cuprate high-temperature superconductors, and which have been used to distinguish the different transport regimes. We observe two distinct features in the $R(T)$ curve: First, a local minimum interpreted as the 3D-2D crossover. Second, a local maximum (indicated with $T_{max}$) around 11.4 K; a signature that has been interpreted as the onset of superconducting fluctuations.

Spatial variations of the spectral gap
Differential conductance maps for both ALD grown and sputtered samples were measured in order to record the spatial variations of the spectral gap of TiN. Fig. S4a and b show an intensity plot and histogram of the gap variations in a 50 nm x 50 nm area at 2.2 K for the ALD-grown sample. The spectral gap is extracted by minimization of the second derivative of the measured differential conductance spectrum with respect to energy at each location. We find a mean gap of 1.78 meV with a standard deviation of 0.63 meV, which corresponds to ~36 % gap variations. The gap variations in our sample are similar to previously reported ones on 50 nm-thick sputtered TiN films *(42)*. Compared to thinner films (3.6 nm, Ref. *(43)*) our gap variations are significantly larger.
Similarly for the sputtered sample (Fig. S4c and d) we extract the gap variations in the same way to find a mean of 0.85 meV and a standard deviation of 0.11 meV corresponding to ~13% gap variations.

Influence of thermal broadening and conventional Andreev reflections on the filling of the spectral gap
In Fig. S5 we show that thermal broadening and conventional Andreev reflection cannot cause the filling of the spectral gap as shown in Fig. 4a. We start with density of states from Dynes formula representing a simple s-wave BCS gap with a fixed size of 1.78 meV for all temperatures,

$$N(E, \Gamma, \Delta) = \left| \text{Re} \frac{E + i\Gamma}{\sqrt{(E + i\Gamma)^2 - \Delta^2}} \right|$$

and simulate integrated differential conductance assuming a constant density of states for the tip

$$g(V, T) = \int_{-\infty}^{+\infty} N(E, \Gamma, \Delta) \left[ -\frac{\partial f(E + eV, T)}{\partial V} \right] t(E) dE,$$

where $t(E)$ is tunneling probability and $f(E,T)$ is the Fermi–Dirac distribution at temperature $T$. The resulting filling of the gap due to thermally excited quasiparticles only contribute to up to 22% of the normal state conductance in the gap at 7.2 K. However, as shown in Fig. S5c, even at 2.2 K the spectral gap is 91% filled.

We further include conventional Andreev reflection in the density of states (equivalent to the $T=0$ K line with two small peaks in Fig. S5b), in the form of a pair of Lorentzian functions centered at ±0.9 meV with an amplitude same as $\max[N(E, \Gamma, \Delta)]$ as an example. As described in the main text, Andreev processes occur with probability $t^2$, while single electron tunneling occurs with probability $t$, where $t$ is the transparency of the barrier. For the extreme case of $t=10^{-2}$, the contribution of Andreev process to in-gap conductance is still negligible at our measuring temperature above 2.2 K. Therefore, we conclude that thermal broadening and conventional Andreev reflections cannot not account for the filling of the spectral gap.

Effect of temperature on the sharpness of the noise enhancement at $V=\Delta$

In the main text we use a simple phenomenological Fermi function as a guide to the eye to the data (blue lines in Fig. 2 and Fig. 3) to include the effect of temperature on the step from $1e$ to $2e$. The reason that the noise does not rise immediately at $\Delta$, but at energies just below $\Delta$ is based on a scattering matrix formalism model that we will describe below for comparison, either model gives similar $S(V)$ curves, as shown in Fig. S8.

We calculate the current and the current noise in a normal metal – insulator – superconductor (NIS) junction using the reflection matrix with electron-hole grading of excitations at energy $\epsilon$, adapted from the scattering matrix formalism of Refs *(44,45)*:

$$r(\epsilon) = \begin{pmatrix} r_{ee}(\epsilon) & r_{eh}(\epsilon) \\ r_{he}(\epsilon) & r_{hh}(\epsilon) \end{pmatrix}.$$

The tunneling barrier is described by a normal scattering matrix for electrons,

$$S = \begin{pmatrix} \rho & \tau \\ \tau & \rho' \end{pmatrix},$$

with the transmission $|\tau|^2 = t$ and $\rho' = -\rho^* \tau/\tau^*$ assumed to be energy independent. Current conservation implies that the scattering matrix $S$ is an unitary matrix ($S^*S = SS^* = 1$), thus $|\tau|^2 + |\rho|^2 = 1$. The corresponding matrix for holes is $S^*$.

The N-S interface is described by the Andreev reflection amplitude $a(\epsilon)$ given by

$$a(\epsilon) = \frac{1}{\Delta} \begin{cases} \epsilon - sng(\epsilon)\sqrt{\epsilon^2 - \Delta^2}, & \text{for } |\epsilon| > \Delta \\ \epsilon - i\sqrt{\Delta^2 - \epsilon^2}, & \text{for } |\epsilon| < \Delta \end{cases}$$

where $\Delta$ is the magnitude of the pair-breaking gap. The reflection amplitudes $r_{ee}(\epsilon)$ and $r_{he}(\epsilon)$ in the scattering matrix can be found from an infinite series expansion for all possible electron and hole trajectories in the junction. By taking the distance between the N-S interface to zero, one can approximate these infinite series by:

$$r_{ee}(\epsilon) = \rho + \frac{\tau^2 \rho'^* a^2(\epsilon)}{1 - |\rho|^2 a^2(\epsilon)}$$

$$r_{he}(\epsilon) = \frac{|\tau|^2 a(\epsilon)}{1 - |\rho|^2 a^2(\epsilon)}$$

In order to calculate the current and current noise we need to define the expectation amplitudes for these reflections

$$R_{ee}(\epsilon) = |r_{ee}(\epsilon)|^2$$
$$= (1-t) - \frac{t(1-t)a^{2*}}{1-(1-t)a^{2*}} - \frac{t(1-t)a^2}{1-(1-t)a^2}$$
$$+ \frac{t^2(1-t)|a|^2}{1-(1-t)a^{2*} - (1-t)a^2 + |a|^2(1-t)^2}$$

and

$$R_{he}(\epsilon) = |r_{he}(\epsilon)|^2 = \frac{t^2|a|^2}{1-(1-t)a^{2*} - (1-t)a^2 + |a|^2(1-t)^2}$$

where we used that $|\tau|^2 = t$ and $|\rho|^2 = 1 - t$ so that we can express these reflection amplitudes in terms of the transmission of the NIS junction $t$. Now the current and the current noise are obtained by

$$I = \frac{2e^2}{h} \int_0^{eV} d\epsilon [1 - R_{ee}(\epsilon) - R_{he}(\epsilon)][f(\epsilon - eV) - f(\epsilon)]$$

where $e$ is the elementary charge, $h$ is Planck's constant, $V$ is the applied bias voltage to the junction and we included the thermal resolution of the experiment by the Fermi-Dirac distribution $f(\epsilon) = \frac{1}{1+e^{\epsilon/k_B T}}$, where $k_B$ is Boltzmann's constant and $T$ is the temperature. The noise is found by

$$S_I = \frac{4e^2}{h} \int_0^{eV} d\epsilon [R_{ee}(\epsilon)[1 - R_{ee}(\epsilon)] + R_{he}(\epsilon)[1 - R_{he}(\epsilon)] + 2R_{ee}(\epsilon)R_{he}(\epsilon)][f(\epsilon - eV) - f(\epsilon)]$$

To simulate the effective charge transferred in the junction we calculate the ratio $S_I/2I$ for each bias voltage $V$. An example fit to the data is shown in Fig. S8.

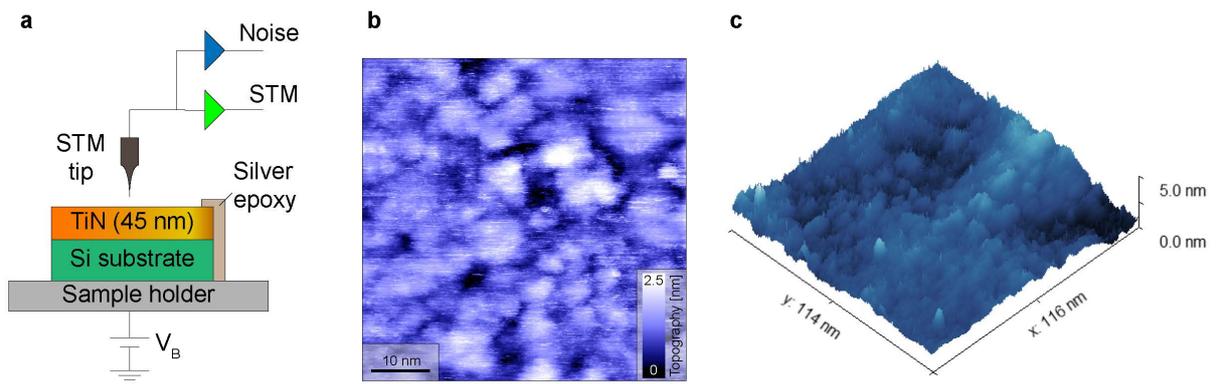

**Fig. S1.**

Measurement schematics and TiN topography. (A). STM measurement schematic (B). Topography of TiN. (C). Three-dimensional view of the TiN topography. The topography in B is obtained from a cropped area of the larger topograph in C. Setup conditions: $V_{bias}$=5 mV, $I_{set}$=100 pA.

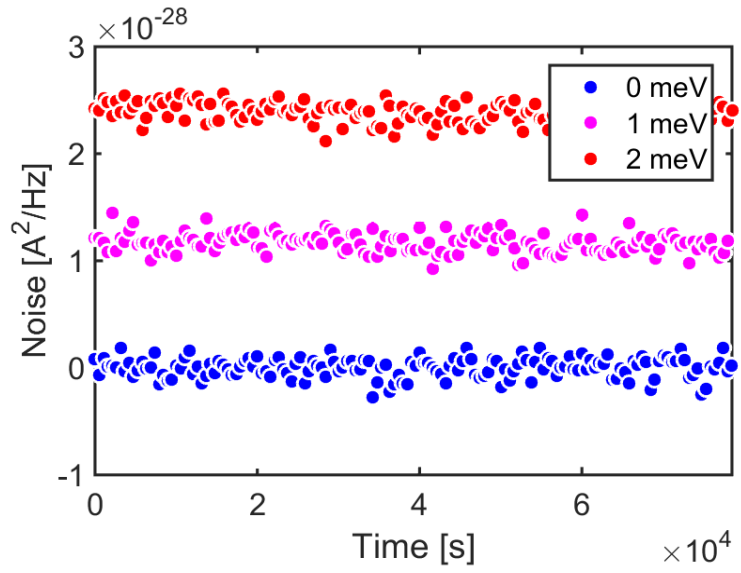

**Fig. S2.**

Noise time traces. Time traces of noise at $R_J$=2.5 MΩ and $T$=2.5 K for 0 meV (with the tip retracted), 1 meV and 2 meV energy. The average standard deviation of these time traces yields an error for the noise figures.

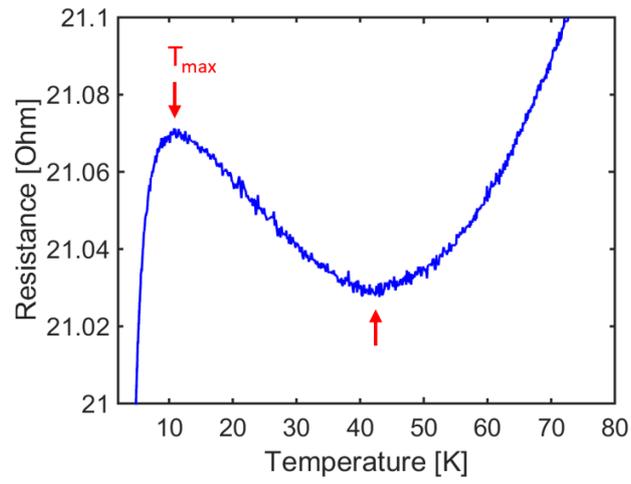

**Fig. S3.**

Detail of $R(T)$ curve. The characteristic N-shaped curvature of $R(T)$ above $T_c$ can be used to distinguish different transport regimes. The local minimum (lower red arrow) indicates the 3D-2D crossover. The local maximum (indicated with $T_{max}$) gives the onset temperature below which transport is governed by superconducting fluctuations.

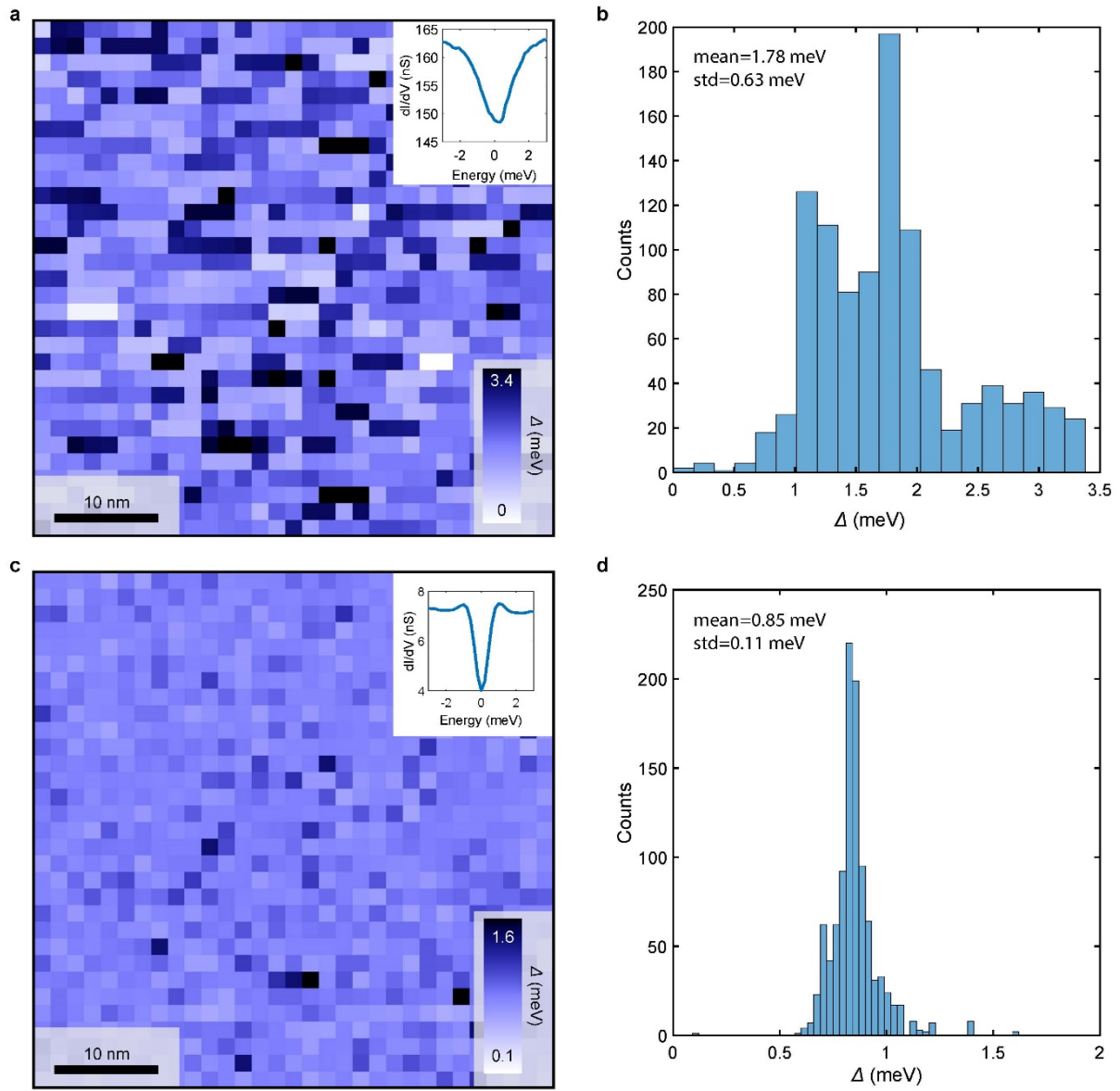

**Fig. S4.**
Mapping the spectral gap of TiN. (A). Map of the spectral gap of TiN (grown by ALD) in a 50 nm x 50 nm area at 2.2 K. Inset: average spectrum of the dI/dV spectra in the area. (B). Histogram of the gap values in a. A mean gap of 1.78 meV is observed and a standard deviation of 0.63 meV. (C). Same as in A for the TiN sample that was grown via sputtering. (D). Same as in b. for the TiN sputtered sample. A mean of 0.85 meV and standard deviation of 0.11 meV is observed.

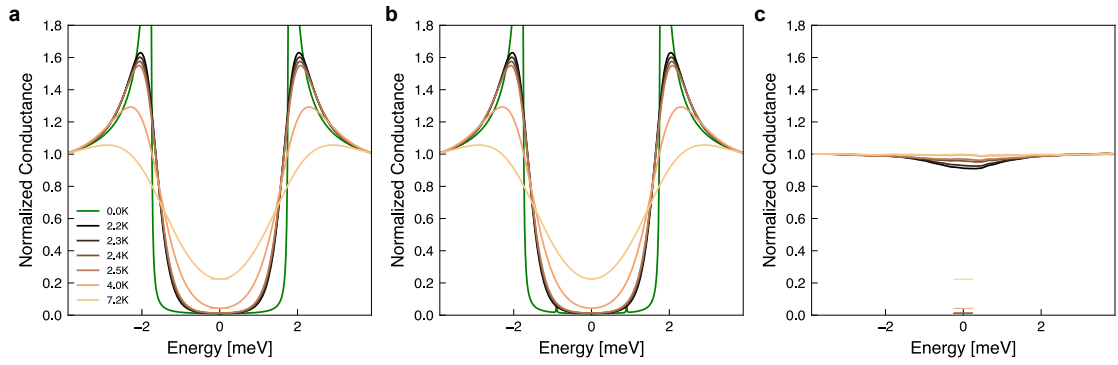

**Fig. S5**
Filling of the spectral gap due to thermal broadening and conventional Andreev reflections. (A). Simulation of normalized differential conductance as a function of energy at various temperatures for an s-wave BCS superconductor with a constant gap size Δ=1.78 meV. (B). same as a except two peaks inside the gap representing Andreev reflections are included in the zero-temperature density of states with barrier transparency of the tunnel junction $t=10^{-2}$. (C). Normalized differential conductance data from Fig. 4a. The horizontal bars show contribution to filling of the spectral gap by thermal broadening and Andreev reflections for different temperatures.

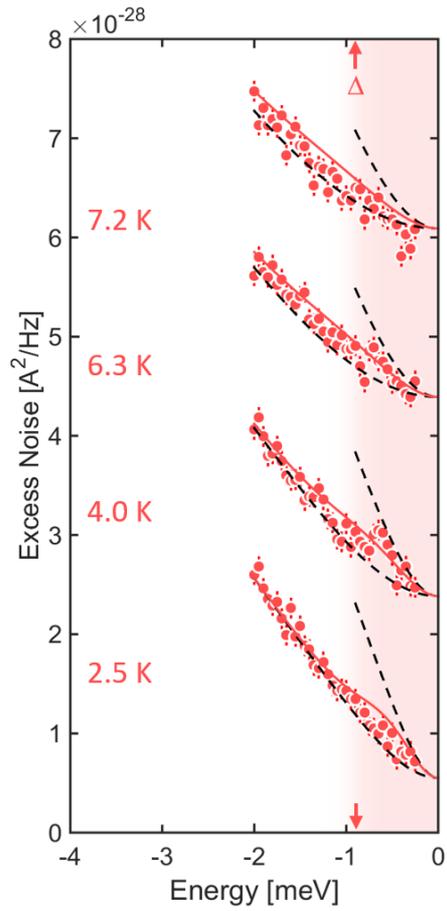

**Fig. S6**

Enhanced noise above Tc for the sputtered sample. Noise spectroscopy on sputtered TiN sample for varying temperatures 2.5 K to 7.2 K. Red dots indicate the measured excess noise in the junction as function of bias voltage at a junction resistance of $R_J$ = 2.5 MOhm. The different temperature curves are offset for clarity. Dashes lines indicate the expected noise for $q^*=1e$ and $q^*=2e$. Red shading highlights the spectral gap measured in the differential conductance.

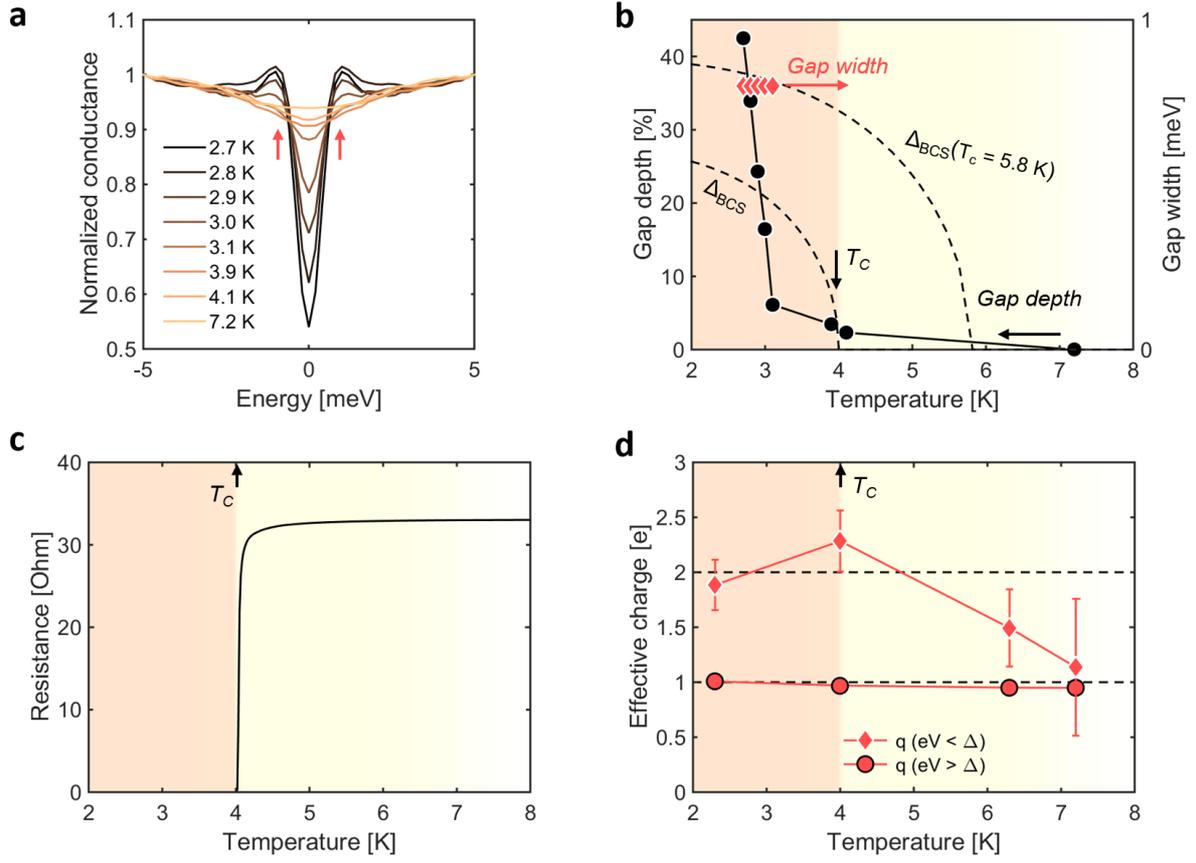

**Fig. S7**

Evidence for a preformed-pair phase above $T_c$ for the sputtered sample. (A). Temperature dependence of the spectral density gap measured by the differential tunneling conductance between 2.7 K and 7.2 K. Red arrows indicate the gap width below $T_c$ determined by minimum of second derivative. Setup conditions: $V_{bias}$ = 2 mV, $I_{set}$ = 200 pA. (B). Gap width (red diamonds) determined via the second derivative of the curves in panel a. The dashed curves indicates the mean-field prediction for $\Delta(T_c = 4\text{ K})$ and $\Delta(T_c = 5.8\text{ K})$ from BCS theory. The depth of the gap at zero bias (black dots) for the curves in panel a is shown in percentages with respect to the conductance at energies outside the gap. (C). Resistance versus temperature curve of our sputtered TiN sample. The orange shaded region indicates the phase-coherent superconducting phase below the transition temperature (D). Effective charge outside (diamonds) and inside (circles) the spectral gap as function of temperature. We define the temperature $T_p$ at the drop of the noise.

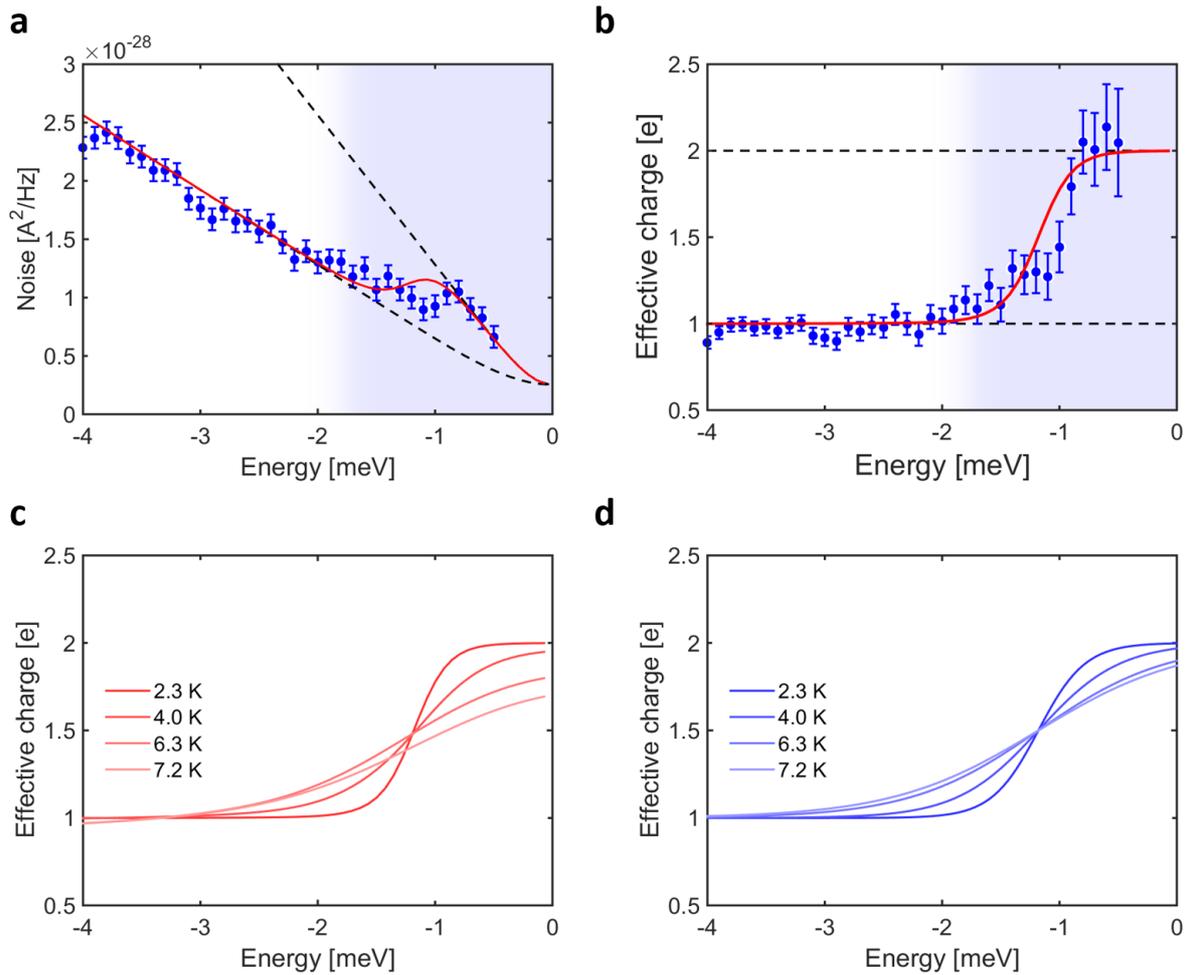

**Fig. S8**

Random-scattering matrix fit to the noise data. (A). The measured noise at 2.3 K as function of the bias voltage (blue dots) and random-scattering matrix simulation fit (red line). Blue shading indicates the spectral gap observed in the differential conductance (B). Similar to a. but now for the effective charge. (C). Random scattering matrix simulation for the four different temperatures used in the experiment. (D). Guides to the eye used in the main text.